\documentclass[%
 amsmath,amssymb,
 aps, physrev,
 twocolumn,
floatfix,
hidelinks
]{revtex4-2}

\usepackage{color}

\usepackage{graphicx}
\usepackage{dcolumn}
\usepackage{bm}
\usepackage{hyperref}

\usepackage{xurl}

\begin{document}

\preprint{APS/123-QED}

\title{\textbf{Brownian yet non-Gaussian diffusion through equilibrium nonlinear friction 
} 
}%

\author{Jakob Mihatsch}
\email{jakob.mihatsch@ovgu.de}
\affiliation{Otto-von-Guericke-Universit{\"a}t Magdeburg, Institut f{\"u}r Physik, Universit{\"a}tsplatz 2, 39106 Magdeburg, Germany}
\author{Andreas M. Menzel}
\email{a.menzel@ovgu.de}
\affiliation{Otto-von-Guericke-Universit{\"a}t Magdeburg, Institut f{\"u}r Physik, Universit{\"a}tsplatz 2, 39106 Magdeburg, Germany}

\date{\today}

\begin{abstract}
In Brownian yet non-Gaussian diffusion (BnGD) the mean squared displacement grows linearly in time. However, the displacement statistics do not follow a normal distribution throughout. 
Typically, they are non-Gaussian at intermediate times, before they cross over to Gaussian in the long-time regime.
We demonstrate that nonlinear friction under correctly applied stochastic equilibrium conditions provides an explanation of this phenomenon also for homogeneous environments. 

\end{abstract}

\maketitle

There has been a growing number of observations of so-called Brownian yet non-Gaussian diffusion (BnGD) in the last two decades \cite{nampoothiri2022brownian}. During such diffusive processes, the mean squared displacement (MSD) grows linearly in time, referred to as Brownian. However, at least at intermediate times, the displacement statistics become non-normal, that is, non-Gaussian. Examples include diffusion of tracers on lipid-bilayer tubes or in actin networks \cite{wang2009anomalous}, in nematic solutions \cite{wang2012brownian}, crowded lipid bilayers \cite{jeon2016protein}, and glass-forming liquids \cite{rusciano2022fickian}, but also in dilute colloidal suspensions \cite{guan2014even} and binary gasses \cite{nakai2023fluctuating}.

The textbook theory of Brownian motion is given by the Langevin equation \cite{gardiner2009stochastic}. It describes a thermally driven particle that is subject to a friction force linear in velocity, for example Stokes' drag.
The resulting displacements are normally distributed according to Gaussian statistics at all times.
However, there are situations where nonlinear friction and/or drag forces apply. For example, starting from microscopic interactions between particles, one can show that the frictional term in a Langevin equation for test particles in a dilute binary gas is complex \cite{ferrari2007particles,ferrari2014particles}. Similarly, starting from a continuum description of a spherical particle in a shear-thinning fluid, one obtains a nonlinear correction to the drag force \cite{datt2018dynamics}.

In the standard Langevin equation, the strength of the stochastic force representing thermal fluctuations is not independent of the strength of the friction force. Instead, in thermal equilibrium, it is dictated by the requirement that stationary distribution of the velocity follows the usual Maxwell-Boltzmann statistics \cite{kubo1966fluctuation}. For linear friction and drag forces, this condition leads to constant strength of the thermal force \cite{menzel2022low}. For the nonlinear Langevin equation, such relations likewise exist, yet generally leading to a nonconstant strength of the stochastic force \cite{klimontovich1994nonlinear}. 

Existing descriptions of stochastic motion under nonlinear friction were shown to exhibit BnGD \cite{goohpattader2010diffusive,menzel2011effect,menzel2015velocity,das2017single,lequy2023stochastic,howlader2025velocity}. However, they often do not adjust the noise strength to obtain equilibrium velocity statistics. Therefore, they assume that the source of the stochastic force is not thermal but given by an external driving force. For instance, this applies to objects under friction on a vibrating substrate \cite{gennes2005brownian,goohpattader2010diffusive}.
In this context, it has been generally remarked that BnGD can be observed in systems of external \textit{non}thermal noise \cite{bialas2020colossal}, implying non-Gaussian velocity distributions \cite{giona2026homogeneous}.

Turning back to equilibrium systems, BnGD is often regarded as a superposition of diffusion processes with different diffusivities \cite{wang2012brownian,metzler2020superstatistics}.
Specifically, we mention in this context
the phenomenological model of ``diffusing diffusivity'' \cite{chubynsky2014diffusing}.
There, the 
diffusion coefficient of a stochastic object is subject to another, independent stochastic process.
For example, it represents a friction force that varies with stochastic motion through a heterogeneous environment of the diffusing particle, or fluctuations of some macroscopic variable.

Here, we return to less sophisticated descriptions. However, we can demonstrate that they are sufficient to describe BnGD. We focus on homogeneous environments and constant conditions for the diffusing particle. Thus, the strength of the friction or drag force, and therefore of the ``diffusivity'' set by the environment, remain \textit{constant}. Yet, friction or drag are nonlinear in velocity. At the same time, we focus on \textit{equilibrium} situations. From the point of view of the moving particle, such situations imply a velocity-dependent strength of thermal noise. 

We show that these ingredients in combination, that is, nonlinear friction and velocity-dependent thermal noise as required in equilibrium, are sufficient to obtain BnGD under equilibrium conditions.
The velocity distribution remains Gaussian and the diffusive process is Brownian (linear increase of the mean squared displacement in time).
Yet, on intermediate time scales, a non-Gaussian displacement distribution (DisD) arises. 

The problem of defining a Langevin equation with a nonlinear friction force has been addressed by Klimontovich over three decades ago \cite{klimontovich1994nonlinear}.
We consider a thermally driven particle, whose position $x$ changes with velocity $\dot{x}=v$. For simplicity, this particle is confined to one spatial dimension \cite{wang2009anomalous}. The probability density of the velocity $p(v,t)$ is governed by the Fokker-Planck equation
\begin{gather}
    \partial_t p=\partial_v\left[\left(\frac{\lambda(v)}{m}v+\frac{B(v)}{m^2}\partial_v\right)p\right].
    \label{eq:FP_kinetic}
\end{gather}
In this equation, $m$ is the mass of the particle and $\lambda(v)$ is a friction or drag function that can generally depend on the velocity.
One can easily verify that for arbitrary $\lambda(v)$ Eq.~\eqref{eq:FP_kinetic} describes an equilibrium system, if the strength of the diffusive term $B(v)$ satisfies
\begin{equation}
    B(v)=\lambda(v)k_BT.
    \label{eq:fluc_diss}
\end{equation}
This means, its stationary solution for the probability is of Gaussian shape that leads to $m\langle v^2\rangle=k_BT$, where $k_B$ is the Boltzmann constant and $T$ marks the temperature.
The Fokker-Planck equation Eq.~\eqref{eq:FP_kinetic} then corresponds to the Itô stochastic differential equation (SDE) \cite{gardiner2009stochastic,hanggi1978stochastic}
\begin{gather}
    m\dot{v}={}-\lambda(v)v+\frac{k_BT}{m}\lambda'(v)+\sqrt{2k_BT\lambda(v)}\eta,
    \label{eq:langevin_ito}
\end{gather}
where $\lambda'$ is the derivative of $\lambda$ with respect to $v$. $\eta$ is a Gaussian white stochastic process with $\langle\eta(t)\rangle=0$ and $\langle\eta(t)\eta(t')\rangle=\delta(t-t')$. 

We note that there are some subtleties in defining a Fokker-Planck equation or corresponding SDE for a given nonlinear friction law. The reasons are multiple possible forms that lead to the same stationary distribution. Equation \eqref{eq:FP_kinetic} is in the so-called kinetic form. As an advantage, it results in the familiar fluctuation-dissipation relation Eq.~\eqref{eq:fluc_diss}. Other forms essentially absorb (partially or completely) the additional drift term in Eq.~\eqref{eq:langevin_ito} into the definition of the friction or drag function $\lambda$ and then express the noise strength in terms of this new friction or drag. For further details, we point to Ref.~\onlinecite{klimontovich1994nonlinear}. 
Here, we just note that any Fokker-Planck equation and corresponding Itô SDE describing a thermally driven particle subject to nonlinear friction can be adapted to the form of Eqs.~\eqref{eq:FP_kinetic} and \eqref{eq:langevin_ito} by redefining $\lambda(v)$.

A prominent example of nonlinear friction is given by so-called Coulomb (or dry) friction \cite{gennes2005brownian,hayakawa2005langevin,touchette2010brownian}, often used to model friction between two solids. Here, $\lambda(v)=\mu\ \mathrm{sgn}(v)/v$, with $\mathrm{sgn}$ denoting the sign function. The magnitude of this type of friction is constant, with absolute value $\mu$, and it is always opposed to the direction of motion of the particle. It corresponds to the situation of an object sliding on a solid substrate. To avoid the jump of the force at $v=0$, the form
\begin{equation}
\lambda(v)=\lambda_0\,\frac{\epsilon}{v}\tanh\!\left(\frac{v}{\epsilon}\right)
\label{eq:tanh}
\end{equation}
smoothly crosses $v=0$. 
The parameter $\epsilon$ has the dimension of velocity. It determines how steep the crossover is at $v=0$. Dry friction follows in the limit $\epsilon\rightarrow 0$. If, instead, $\epsilon\gg \sqrt{k_BT/m}$, there is only slight deviation from the linear case $\lambda(v)=\lambda_0$ around $v=0$.

In general, friction that increases sublinearly with $v$ can imply that forces exerted by the particle decrease friction with the environment. This is a characteristic of shear thinning \cite{lequy2023stochastic}. Stokes drag on a sphere immersed in a shear-thinning fluid has been shown to increase nonlinearly with velocity \cite{datt2018dynamics}.

For comparison, we also consider a drag that increases stronger than linearly with velocity,
\begin{equation}
    \lambda(v)=\lambda_0(1+\alpha v^2),\quad \alpha>0.
    \label{eq:larger}
\end{equation}
Here, $\alpha$ is a parameter of dimension $m/k_BT$.
This form
can be considered as a first-order Taylor expansion of any more complicated friction force.
An analytical result for the diffusion coefficient of a particle described by the corresponding nonlinear Langevin equation is available \cite{lindner2007diffusion}.
A Langevin equation with a similar nonlinear amendment
was explicitly derived theoretically for a dilute-gas scenario
\cite{ferrari2007particles,ferrari2014particles} and verified experimentally \cite{hohmann2017individual}.
Similarly to what was discussed above, a super-linearly increasing friction can be interpreted as motion in a shear-thickening medium \cite{menzel2015velocity}.

We integrate Eq.~\eqref{eq:langevin_ito} numerically using a Euler-Maruyama scheme \cite{Kloeden2010oq}.
As a measure of non-Gaussianity of the DisD, we choose the excess kurtosis
\begin{equation}
    \gamma= \frac{\langle x(t)^4\rangle}{\langle x(t)^2\rangle^2}-3.
    \label{eq:kurtosis}
\end{equation}
To test linearity of the MSD in time, we use the slope of the MSD $\mathrm{d}\langle{x^2}\rangle/\mathrm{d}t$.
The initial position is $x=0$, and initial velocities are drawn from the stationary distribution, a Gaussian with $m\langle v^2\rangle=k_BT$.

\begin{figure*}
        \centering
        \includegraphics[width=\linewidth]{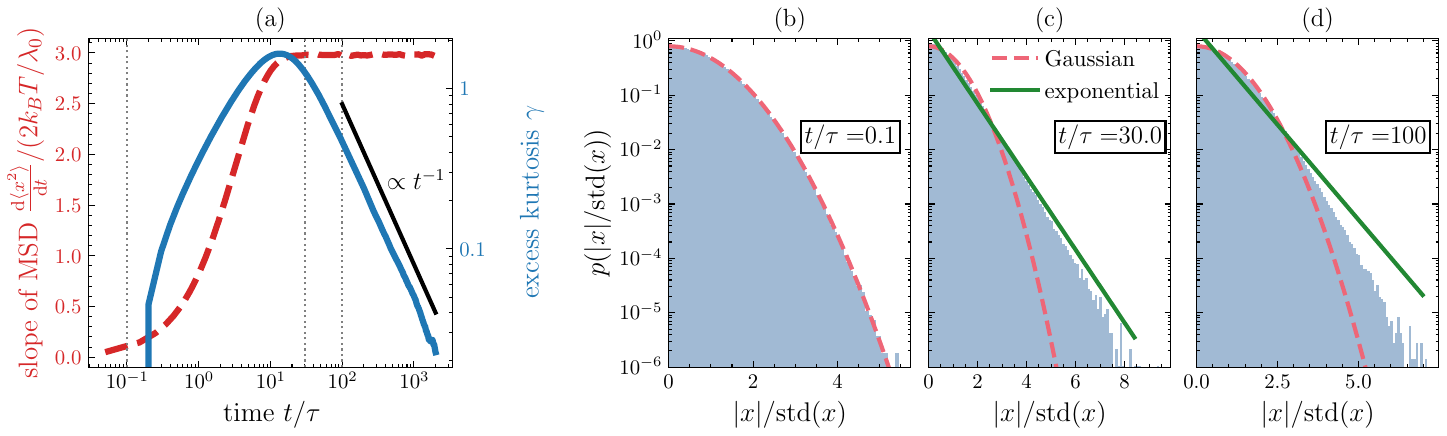}
        \caption{Numerical results obtained from integrating Eq.~\eqref{eq:langevin_ito} with $\lambda(v)=\lambda_0\epsilon\tanh(v/\epsilon)/v$ and $\epsilon^2=0.1k_BT/m$, using an Euler-Maruyama scheme. (a) Evolution of the slope of the MSD $\mathrm{d}\langle{x^2}\rangle/\mathrm{d}t$ (red line) and the excess kurtosis $\gamma$  (blue line) over time in units of $\tau=m/\lambda_0$. The black line reflects a power-law dependence $\propto t^{-1}$ included for comparison.
        (b)--(d) Histograms of the displacement distributions $p(|x|/\mathrm{std}(x))$ as a function of overall displacements $x$ normalized by their standard deviation $\mathrm{std}(x)$ at the times indicated by vertical dotted lines in (a), in chronological order. Pink dashed lines are Gaussian fits to the data, the green solid line represents an exponential distribution.}
        \label{fig:tanh}
    \end{figure*}

Figure \ref{fig:tanh} shows the result for the case of $\lambda(v)=\lambda_0\epsilon\tanh(v/\epsilon)/v$ with $\epsilon^2=0.1k_BT/m$. The simulations were performed with timestep $\delta t=0.1m/\lambda_0$ and for $10^7$ realizations.
At very small times, the displacements follow a Gaussian distribution. 
The MSD initially develops through typical ballistic behavior, before it crosses over into the diffusive regime of constant slope and thus linear increase in time.
During the initial period, the displacement distribution becomes less Gaussian and the excess kurtosis grows. It reaches a maximum at approximately the crossover of the MSD from ballistic to diffusive. At this point, the displacement distribution can be closely approximated by an exponential distribution. Importantly, from this point onwards, non-Gaussianity persists and only gradually decreases, while the MSD continues to grow linearly with constant diffusivity. The excess kurtosis decays approximately according to a power law $\gamma\propto t^{-1}$. 

\begin{figure*}
        \centering
        \includegraphics[width=\linewidth]{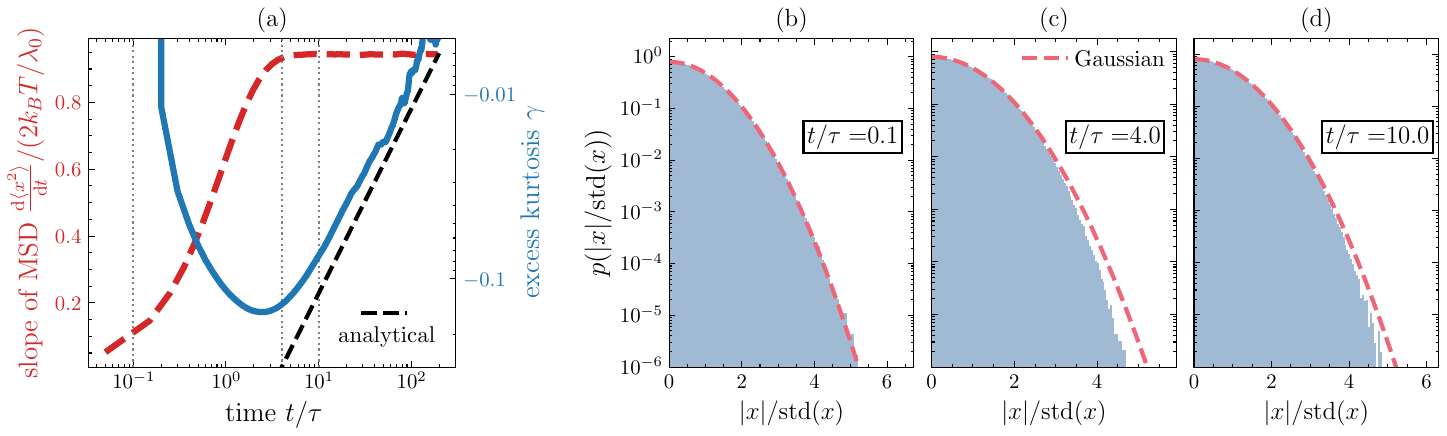}
        \caption{Numerical results of integrating Eq.~\eqref{eq:langevin_ito} with $\lambda(v)=\lambda_0(1+\alpha v^2)$ and $\alpha=0.05m/(k_BT)$, using an Euler-Maruyama scheme. (a) Evolution of the slope of the MSD $\mathrm{d}\langle{x^2}\rangle/\mathrm{d}t$ (dashed red line) and the excess kurtosis $\gamma$  (solid blue line) over time in units of $\tau=m/\lambda_0$. The dashed black line is the analytical prediction from Eq.~\eqref{eq:kurtosis_result}. 
        (b)--(d) Histograms of the displacement distributions $p(|x|/\mathrm{std}(x))$ as a function of overall displacements $x$ normalized by their standard deviation $\mathrm{std}(x)$ at the times indicated by vertical dotted lines in (a), in chronological order. Dashed pink lines correspond to Gaussian fits to the data.}
        \label{fig:larger}
    \end{figure*}

We perform analogous simulations for the second case, $\lambda(v)=\lambda_0(1+\alpha v^2)$, with $\alpha=0.05m/(k_BT)$, timestep $\delta t=0.1m/\lambda_0$, and $2\times10^7$ realizations. The results are visualized in Fig.~\ref{fig:larger}.
There, similarly, after an initial ballistic regime, non-Gaussian displacement statistics coexist with a linear increase of the MSD in time. We observe the same approximate power-law decay of the kurtosis.
Here, friction grows more strongly than linearly for large velocities, in contrast to the situation in Fig.~\ref{fig:tanh}.
This leads to a narrower DisD than Gaussian, in contrast to the broader one in Fig.~\ref{fig:tanh}. The tails of the distribution decay faster than Gaussian and the excess kurtosis becomes negative, see Fig.~\ref{fig:larger}.

Intuitively, the described observations can be viewed as follows. 
In equilibrium, the overall stationary velocity distribution always remains of Gaussian shape. 
However, the temporal evolution of the velocity for a given particle is different from the case of linear friction. If the friction force increases sublinearly with velocity, the strength of the thermal noise, on average, decreases with increasing velocity. A particle that currently has a large velocity will, on average, keep this velocity for a longer time than a particle subject to linear friction. Similarly, if friction increases superlinearly, the strength of the thermal noise, on average, increases with velocity. Then, large velocities, on average, persist for a shorter time compared to the case of linear friction. This explains why, at intermediate timescales, large displacements can occur with a different probability distribution of longer and shorter tails, respectively, relative to the Gaussian distribution known from linear Brownian motion.

To provide an analytical argument for the emergence of BnGD, we derive an analogue to the diffusion equation for nonlinear Brownan motion. 
We first focus on a drag coefficient $\lambda(v)$ of the form of Eq.~\eqref{eq:larger}, where we can expand the results perturbatively for a small nonlinearity parameter $\bar{\alpha}=\alpha k_BT/m\ll 1$.
We consider motion on a timescale $\tau_D$
much larger than the inertial timescale $\tau=m/\lambda_0$. Thus, we write $\tau_D=\mu^{-1}\tau$, where $\mu\ll 1$ is a small dimensionless quantity. A characteristic lengthscale of diffusion can be defined as $\xi=\sqrt{k_BT\tau_D/\lambda_0}$.
We introduce dimensionless time $\bar{t}$, position $\bar{x}$, and velocity $\bar{v}$ according to
\begin{equation}
    t=\tau_D \bar{t},\quad x=\xi\bar{x},\quad v=\frac{\xi}{\tau_D}\bar{v}.
\end{equation}
The joint probability density $\bar{P}(\bar{x},\bar{v},\bar{t})$ of position and velocity is governed by the Fokker-Planck equation
\begin{multline}
    \partial_{\bar{t}} \bar{P}={}-\partial_{\bar{x}}(\bar{v}\bar{P})\\+\partial_{\bar{v}}\left[\mu^{-1}\left(1+\bar{\alpha}\mu \bar{v}^2\right)\bar{v}+\mu^{-2}\left(1+\bar{\alpha}\mu \bar{v}^2\right)\partial_{\bar{v}}\right]\bar{P}.
    \label{eq:FP_combined}
\end{multline}
To eliminate the velocity from this equation, we employ the method of moments \cite{wilemski1976derivation}, following closely the procedure outlined in Ref.~\onlinecite{goldobin2025kinetic}.
By multiplying Eq.~\eqref{eq:FP_combined} with powers of $\bar{v}$ and integrating, we obtain an infinite hierarchy of equations for the moments $w_n(\bar{x},\bar{t})=\int_{-\infty}^{\infty}\mathrm{d}\bar{v}\,\bar{v}^n\bar{P}(\bar{x},\bar{v},\bar{t}).$
This is the starting point for a systematic expansion in $\mu$ that allows us to obtain a closed equation for the probability density in position space, $w_0(\bar{x},\bar{t})$.
We find
\begin{gather}
    \partial_{\bar{t}} w_0=D\partial_{\bar{x}}^2w_0-D^{(4)}\partial_{\bar{x}}^4w_0+\mathcal{O}(\mu^2,\bar{\alpha}^2),\label{eq:expansion}\\
     D=\sqrt{\pi}\ \sqrt{\frac{1}{2\bar{\alpha}}}\exp\left(\frac{1}{2\bar{\alpha}}\right)\mathrm{erfc}\sqrt{\frac{1}{2\bar{\alpha}}},\\
    D^{(4)}=4\mu\bar{\alpha}+\mathcal{O}(\mu^2,\bar{\alpha}^2).
    \label{eq:D44}
\end{gather}
On the right-hand side of Eq.~\eqref{eq:expansion}, the first term 
corresponds to familiar diffusion and is of zeroth order in $\mu$.
The expression for the diffusion coefficient $D$ agrees with the one found by Lindner \cite{lindner2007diffusion}.
Next, the term linear in $\mu$, see Eq.~\eqref{eq:D44}, includes fourth-order derivatives with respect to $\bar{x}$. 
For linear friction ($\alpha=0$), 
we recover previous results \cite{wilemski1976derivation,gardiner2009stochastic}.

\begin{figure}
        \centering
        \includegraphics[width=\linewidth]{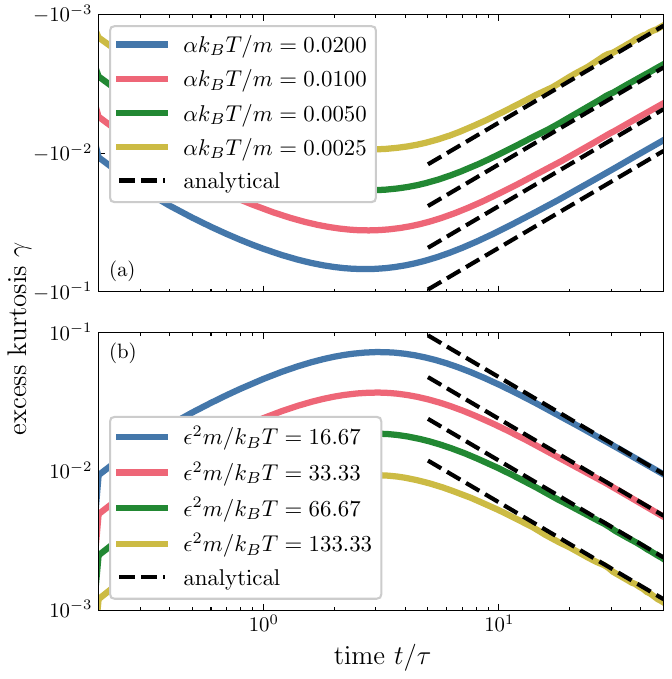}
        \caption{Excess kurtosis $\gamma$ evaluated from  numerical simulations according to Eq.~\eqref{eq:langevin_ito} as a function of time $t$ for different strengths of nonlinear friction. Analytical predictions are included as black dashed lines following Eq.~\eqref{eq:kurtosis_result}. (a) Results for the friction force according to Eq.~\eqref{eq:larger} and varying values of the rescaled friction parameter $\bar{\alpha}=\alpha k_BT/m$.
        (b) Analogous results for the $\tanh$-friction introduced in Eq.~\eqref{eq:tanh} and varying values of its strength as set by the parameter $\epsilon$. For comparison with the analytical prediction in Eq.~\eqref{eq:kurtosis_result} and the results in (a), we
        use the relation $\alpha=-1/(3\epsilon^2)$.}
        \label{fig:complex}
    \end{figure}

The higher-order spatial derivatives in Eq.~\eqref{eq:expansion} are connected to non-Gaussian displacement statistics.
By multiplying Eq.~\eqref{eq:expansion} with $\bar{x}^2$ and integrating, we find that the MSD still increases linearly
\begin{equation}
    \langle \bar{x}^2(\bar{t})\rangle=2D\bar{t}
\end{equation}
Proceeding analogously for $\bar{x}^4$, we find an expression for the excess kurtosis
\begin{equation}
    \gamma={}-6\frac{D^{(4)}}{D^2}\bar{t}^{-1}={}-24\bar{\alpha}\mu \bar{t}^{-1}+\mathcal{O}(\mu^2,\bar{\alpha}^2).
    \label{eq:kurtosis_result}
\end{equation}
In Fig.~\ref{fig:complex}(a), we compare the kurtosis predicted by Eq.~\eqref{eq:kurtosis_result} with simulation results for varying values of $\bar{\alpha}=\alpha k_BT/m$, based on a timestep $\delta t=0.1m/\lambda_0$ and $10^{10}$ realizations. Theoretical precitions and numerical results agree well for sufficiently small parameters $\bar{\alpha}$ quantifying nonlinearity.

Finally, we return to the case of friction that increases more slowly than linearly with velocity. In Eq.~\eqref{eq:larger}, this would imply 
$\alpha<0$.
Strictly speaking, $\alpha<0$ for velocities $v^2>-\alpha$ leads to a ``frictional'' force that acts into the direction of motion.
Thus, it adds energy and therefore destabilizes the system.
However, we can view this case as a first-order Taylor expansion of the tanh-friction in Eq.~\eqref{eq:tanh} and consider sufficiently small values of $\alpha$. It leads to $\alpha=-1/(3\epsilon^2)$ when we compare with the predictions of Eq.~\eqref{eq:kurtosis_result}.

For this case, Fig.~\ref{fig:complex}(b) shows simulation results based on the $\tanh$-friction in Eq.~\eqref{eq:tanh}. The analytical results from the lowest-order expansion in the nonlinearity parameter $\bar{\alpha}$ is quantitatively accurate even for larger values of $|\bar{\alpha}|$ than in Fig.~\ref{fig:complex}(a). The $\tanh$-form of the friction function for larger velocities $|v|$ strictly limits frictional values, in contrast to the unbounded growth for the quadratic form in Fig.~\ref{fig:complex}(a).

In summary, we have investigated the one-dimensional stochastic motion of a particle subject to nonlinear friction in equilibrium with its environment.
The nonlinearity of the friction force leads to non-Gaussian displacement distributions (DisD), while the mean-squared displacement increases linearly in time.
After initialization, the DisD at very small timescales is Gaussian.
However, afterwards and as a consequence of nonlinear friction, non-Gaussian distributions of the displacements arise at intermediate times. Illustratively, for a friction force that increases more slowly (faster) than linearly with velocity, larger velocities persist longer (shorter) than smaller velocities. This is reflected by corresponding non-Gaussian displacement distributions.
The non-Gaussianity of the DisD is relatively long-lived, as the excess kurtosis decays according to a power law $\gamma\propto t^{-1}$.

If the friction force increases sublinearly with the velocity, the DisD appears to exhibit long exponential tails.
If, instead, the drag increases more strongly than linearly, the DisD becomes narrower than Gaussian.
Interestingly, such a platykurtic DisD has
so far only rarely been predicted, to our knowledge in systems with active particles or externally imposed flows \cite{yin2021non}.

Eliminating velocity from the Fokker-Planck equation using the method of moments, we obtain a diffusion equation.
An expansion for small nonlinearities leads to
higher-order spatial derivatives in the diffusion equation. We showed by analytical calculation that already the first higher-order derivative causes non-Gaussian DisD.

Overall, we have demonstrated that basic equilibrium systems in homogeneous environments feature non-Gaussian displacement statistics in combination with Brownian diffusion, meaning a mean squared displacement that increases linearly in time. The key is a nonlinear friction that triggers, by equilibrium conditions, stochastic forces of temporally varying strength. As such, the setting of nonlinear friction under equilibrium conditions provides a very generic mechanism for the phenomenon of BnGD.

\bibliography{cite}

\end{document}